\documentclass[
    aps,
    pra,
    reprint,
    superscriptaddress,
    showpacs,
]{revtex4-1}

\usepackage{tgtermes}
\usepackage[T1]{fontenc}
\usepackage{amsmath}
\usepackage{amssymb}
\usepackage{bm}
\usepackage{graphicx}
\usepackage[cmyk]{xcolor}
\usepackage{hyperref}
\hypersetup{
    unicode,
    bookmarksnumbered = true,
    bookmarksopen     = true,
    breaklinks        = false,
    colorlinks        = true,
    linkcolor         = blue,
    citecolor         = blue,
    filecolor         = blue,
    urlcolor          = blue,
}
\allowdisplaybreaks[3]

\makeatletter
    \def\DisplaySkip{11pt \@plus 3pt \@minus 1pt}
    \def\VerticalSpaceBeforeSection{16pt \@plus 1ex \@minus 0.2ex}
    \def\VerticalSpaceAfterSection{5pt}
    \g@addto@macro\normalsize{
        \setlength\abovedisplayskip{\DisplaySkip}
        \setlength\abovedisplayshortskip{\DisplaySkip}
        \setlength\belowdisplayskip{\DisplaySkip}
        \setlength\belowdisplayshortskip{\DisplaySkip}
    }
    \def\section{
        \@startsection
        {section}
        {1}
        {\z@}
        {\VerticalSpaceBeforeSection}
        {\VerticalSpaceAfterSection}
        {\normalfont\small\bfseries\centering}%
    }
    \def\subsection{
        \@startsection
        {subsection}
        {2}
        {\z@}
        {\VerticalSpaceBeforeSection}
        {\VerticalSpaceAfterSection}
        {\normalfont\small\bfseries\centering}%
    }
    \def\subsubsection{
        \@startsection
        {subsubsection}
        {3}
        {\z@}
        {\VerticalSpaceBeforeSection}
        {\VerticalSpaceAfterSection}
        {\normalfont\small\itshape\centering}%
    }
\makeatother
\usepackage[truedimen,letterpaper,margin=0.7in,bottom=1in]{geometry}

\providecommand\Parens[1]{\left(#1\right)}
\providecommand\Squares[1]{\left[#1\right]}
\providecommand\braces[1]{\{#1\}}

\providecommand\moment[1]{\langle#1\rangle}

\providecommand\ket[1]{\lvert#1\rangle}
\providecommand\Ket[1]{\left\lvert#1\right\rangle}
\providecommand\bra[1]{\langle#1\rvert}
\providecommand\Bra[1]{\left\langle#1\right\rvert}
\providecommand\braket[3]{\langle#1\rvert#2\lvert#3\rangle}

\providecommand\Abs[1]{\left\lvert#1\right\rvert}

\providecommand\re{\operatorname{Re}}
\providecommand\im{\operatorname{Im}}

\newcommand\psiA{\psi_\mathrm{A}}
\newcommand\WS  {W_\mathrm{S}}
\newcommand\WM  {W_\mathrm{M}}
\newcommand\WA  {W_\mathrm{A}}
\newcommand\Wout{W_\text{out}}

\newcommand\UT{Department of Applied Physics, School of Engineering, The University of Tokyo, 7-3-1 Hongo, Bunkyo-ku, Tokyo 113-8656, Japan}
\newcommand\UP{Department of Optics, Palack\'y University, 17. listopadu 1192/12, 77146 Olomouc, Czech Republic}
\newcommand\UNSW{Centre for Quantum Computation and Communication Technology, School of Engineering and Information Technology, University of New South Wales Canberra, ACT 2610, Australia}

\begin{document}

\title{Implementation of a quantum cubic gate by adaptive non-Gaussian measurement}

\author{Kazunori Miyata}
\email{miyata@alice.t.u-tokyo.ac.jp}
\affiliation{\UT}
\author{Hisashi Ogawa}
\affiliation{\UT}
\author{Petr Marek}
\affiliation{\UP}
\author{Radim Filip}
\affiliation{\UP}
\author{Hidehiro Yonezawa}
\affiliation{\UNSW}
\author{Jun-ichi Yoshikawa}
\affiliation{\UT}
\author{Akira Furusawa}
\email{akiraf@ap.t.u-tokyo.ac.jp}
\affiliation{\UT}

\date{\today}

\begin{abstract}
We present a concept of non-Gaussian measurement composed of a non-Gaussian ancillary state, linear optics and adaptive heterodyne measurement, and on the basis of this we also propose a simple scheme of implementing a quantum cubic gate on a traveling light beam.
In analysis of the cubic gate in the Heisenberg representation, we find that nonlinearity of the gate is independent from nonclassicality; the nonlinearity is generated solely by a classical nonlinear adaptive control in a measurement-and-feedforward process while the nonclassicality is attached by the non-Gaussian ancilla that suppresses excess noise in the output.
By exploiting the noise term as a figure of merit, we consider the optimum non-Gaussian ancilla that can be prepared within reach of current technologies and discuss performance of the gate.
It is a crucial step towards experimental implementation of the quantum cubic gate.
\end{abstract}


\maketitle

\section{Introduction}

Development and application of quantum physics crucially rely on progress in quantum operations with various physical systems.
For discrete-variable systems, a basic controlled-NOT nonlinear gate \cite{Nielsen2000} has been already demonstrated with many systems \cite{OBrien2003, Schmidt-Kaler2003, Plantenberg2007, Zu2014} and the current problem is scalability of their implementations.
On the other hand, for more complex continuous-variable (CV) systems \cite{Furusawa2011}, a full set of basic operations has not been closed yet.
It was proven that in order to synthesize an arbitrary unitary operation, it is enough to add a cubic nonlinear operation to the already existing Gaussian operations \cite{Lloyd1999}.
Any nonlinearity can be principally obtained from a chain of the Gaussian operations, the cubic nonlinearity and feedforward corrections \cite{Lloyd1999, Sefi2011}.
The cubic nonlinearity is therefore a bottleneck of CV quantum physics.

Already a decade ago, Gottesman, Kitaev and Preskill (GKP) suggested a way how to implement a cubic nonlinear gate based on Gaussian operations, Gaussian measurement, quadratic feedforward correction and an ancillary cubic state produced by the cubic nonlinearity \cite{Gottesman2001}.
Various approaches towards the cubic gate have followed \cite{Bartlett2002, Ghose2007, Sasaki2004, Marshall2015}.
Particularly in the field of quantum optics, most of the components of the cubic gate have been experimentally demonstrated, mainly because of the high quality of generating squeezed states and efficient homodyne detection.
The Gaussian operations have been already mastered \cite{Yoshikawa2007, Yoshikawa2008, Yoshikawa2011}, utilizing a concept of measurement-induced operations \cite{Filip2005}.
Furthermore, they have been tested on non-Gaussian states of light \cite{Miwa2014}, to prove their general applicability.
Recently, the quadratic electrooptical feedforward control has been demonstrated \cite{Miyata2014}.
In addition, to independently obtain the cubic state, a finite dimensional approximation of the cubic state has been suggested \cite{Marek2011} and its performance in the GKP scheme has been discussed.
The cubic state has been experimentally generated as a superposition of photons and verified \cite{Yukawa2013}.
Potentially, such a superposition state can be stored in and retrieved from recently-developed optical quantum memories \cite{Yoshikawa2013,Bimbard2014}.
In order to make resource nonclassical states compatible with the measurement-based scheme, real-time quadrature measurement of a single-photon state has been demonstrated \cite{Ogawa2015}.

A drawback of the original GKP idea is that requires to implement the quantum nondemolition gate, i.e., the CV controlled-NOT gate \cite{Filip2005}, and a squeezing feedforward that depends on the measurement result.
While each of them has been already demonstrated \cite{Yoshikawa2008, Miyata2014}, the total implementation to build a unitary cubic operation demands three squeezed states as well as one non-Gaussian ancilla, and is probably not the simplest arrangement.
In contrast, we here use adaptability of linear optical schemes and propose a better and simpler topology with linear optics and suitable ancillary states.

Our approach is to tuck all the non-Gaussian aspects into the measurement process.
The topology will be then similar to the simple one used for a measurement-induced squeezing gate \cite{Yoshikawa2007, Filip2005, Miwa2009, Miwa2014, Miyata2014}.
Non-Gaussian operations can be realized by simply substituting a measurement of nonlinear combination of quadrature amplitudes for the Gaussian homodyne measurement \cite{Menicucci2006, Gu2009}.
We construct such a measurement in a form of a generalized non-Gaussian measurement by combining ordinary Gaussian measurement tools with non-Gaussian ancillary states that can be prepared with photon detection.
In fact, we can exploit arbitrary superpositions of photon-number states up to certain photon level within reach of current technologies \cite{Bimbard2010, Yukawa2013, Yukawa2013a}.

In this paper, we first provide an idea of non-Gaussian measurement comprising a non-Gaussian ancillary state, linear optics and adaptive heterodyne measurement.
Using the non-Gaussian measurement, we next propose a simple schematic of a quantum cubic gate based on the measurement-induced operation scheme, whose resource states are only one squeezed vacuum and one non-Gaussian state.
While in previous work the input-output relation of the cubic gate has been investigated in the Schr\"odinger picture, here we analyze the gate in the Heisenberg picture to include imperfections in the scheme.
We then find that nonlinearity of the gate is independent from nonclassicality.
Specifically, the nonlinearity is generated solely by a classical nonlinear adaptive control in a measurement-and-feedforward process regardless of the non-Gaussian ancilla.
On the other hand, the nonclassicality is attached by the ancilla that compensates residual noise in the output.
Finally, we discuss an overall performance of the cubic gate in such a topology and consider non-Gaussian ancillary superposition states up to certain photon level to investigate how well the unwanted noise can be suppressed in the gate.

\section{Minimal implementation of measurement-induced quantum operations}
\label{SECMinimalImplementationOfMeasurementInducedQuantumOperations}

Measurement-induced quantum operation scheme \cite{Filip2005} decomposes various quadratic operations into linear optics, displacement operation, homodyne detection and offline squeezed light beams, which are readily available in actual optical experiments.
One of the realizations of the scheme is the basic squeezing gate.
Firstly we combine an input state $\ket{\psi}$ and an eigenstate $\ket{x = 0}$ of the position quadrature $\hat{x}$ at a beam splitter whose transmittance is represented by $\sqrt{T}$.
We then measure the momentum quadrature $\hat{p}$ of one of the optical modes and obtain a value $y$.
Finally we apply displacement to the $p$ quadrature of the remaining mode with the value $p_\mathrm{disp} = \sqrt{(1-T)/T} y$ and obtain a squeezed output state.
Ideally the output is a pure state $\hat{S} \ket{\psi}$ where $\hat{S}$ is an $x$-squeezing operator defined as $\hat{S}^\dagger \hat{x} \hat{S} = \sqrt{T} \hat{x}$ and $\hat{S}^\dagger \hat{p} \hat{S} = \hat{p} / \sqrt{T}$.
In the case of implementing $p$ squeezing, it is enough to replace the ancillary $x$ eigenstate with the $p$ eigenstate $\ket{p = 0}$ and exchange the roles of $x$ and $p$ quadratures.
This type of operation has been successfully demonstrated in \cite{Yoshikawa2007, Miwa2014}, where the position eigenstate is replaced with the squeezed vacuum.

\begin{figure}
\centering
\includegraphics{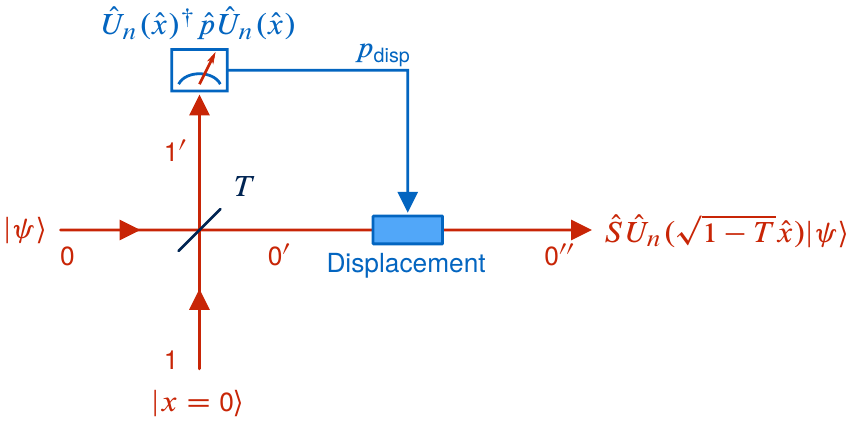}
\caption{(Color online)
Minimal single-mode implementation of measurement-induced quantum operation.
}
\label{FIGMinimalMBQCSchematic}
\end{figure}
On the basis of one-way CV cluster computation \cite{Menicucci2006, Gu2009}, we can generalize the basic squeezing gate to minimal single-mode implementation of arbitrary-order quantum operations as shown in Fig.~\ref{FIGMinimalMBQCSchematic}\@.
The homodyne detector in the squeezing gate is now replaced with a detector that measures a general quadrature $\hat{U}_n^\dagger(\hat{x}) \hat{p} \hat{U}_n(\hat{x})$, where the unitary operator $\hat{U}_n(\hat{x})$ is defined as $n$th-order \emph{phase gate} $\hat{U}_n(\hat{x}) = \exp(i \gamma \hat{x}^n)$ with a real parameter $\gamma$.
Hereafter we set $\hbar = 1$ for simplicity.
The measured general quadrature is thus $\hat{p} + n \gamma \hat{x}^{n - 1}$.
In the ideal case, the output state is expressed as $\hat{S} \hat{U}_n(\sqrt{1 - T} \hat{x}) \ket{\psi}$.
This gate deterministically applies the phase gate to the input state with the additional constant squeezing that can be compensated by another squeezer.

It is known that, arbitrary single-mode unitary can be decomposed into the set of gates $\hat{U}_n(\hat{x})$ for $n = 1, 2, 3$  for all $\gamma \in \mathbb{R}$, together with the $\pi / 2$ phase shift \cite{Gu2009, Sefi2011}.
This also holds when we exploit the minimal implementation in Fig.~\ref{FIGMinimalMBQCSchematic}\@.
$\hat{U}_1(\hat{x})$ is the trivial displacement operation, and $\hat{U}_2(\hat{x})$ has been experimentally demonstrated \cite{Miwa2009, Miyata2014}.
The remained task is thus to realize a cubic gate $\hat{U}_3(\hat{x})$.
We now consider how to construct measurement of the nonlinear quadrature $\hat{p} + 3 \gamma \hat{x}^2$ with affordable apparatuses, as explained in the following sections.

\section{Non-Gaussian Measurement by generalized heterodyne detection}
\subsection{Projecting on pure states}

In quantum physics, measurements are represented by operators.
In the simplest case of von Neumann measurements, these operators are simply projectors on particular quantum states.
In the case of the keystone measurement of CV quantum optics, the homodyne detection, each measurement result indicates the measured state was projected on an eigenstate of the measured quadrature operator.
Analogously, the heterodyne detection, which can be modeled by a pair of homodyne detectors simultaneously measuring conjugate quadratures of a mode split by a balanced beam splitter \cite{Weedbrook2012}, implements a projection onto a coherent state.
Both of these kinds of measurements are Gaussian---the measured quadrature distribution is Gaussian if the measured state is Gaussian.

One way to achieve a non-Gaussian measurement is to take advantage of non-Gaussian states in combination with the standard heterodyne detection schemes.
The basic idea of the measurement is best explained in the $x$ representation.
Consider we have a standard heterodyne detection configuration, where the idle port of the beam splitter is not injected by a vacuum but by a specifically prepared ancillary state $\ket{\psiA}=\int\psiA(x)\ket{x}\,dx$.
For a particular pair of measurement results $q$ and $y$, the procedure implements projection onto a state
\begin{align}
\label{EQNPureHeterodyneProjector}
\hat{D}(\sqrt{2} q + i \sqrt{2} y)\hat{\mathcal{T}}\ket{\psiA}.
\end{align}
Here $\hat{D}(\alpha)=\exp\braces{i \sqrt{2} \hat{x} \im[\alpha] - i \sqrt{2} \hat{p} \re[\alpha]}$ stands for the displacement operator and $\hat{\mathcal{T}}$ is the time-reversal antiunitary operator represented by $\hat{\mathcal{T}}^\dagger \hat{x} \hat{\mathcal{T}} = \hat{x}$ and $\hat{\mathcal{T}}^\dagger \hat{p} \hat{\mathcal{T}} = -\hat{p}$.
To derive the expression~\eqref{EQNPureHeterodyneProjector}, we can start with the projection states of the pair of homodyne detectors
\begin{align}
\bra{x_1=q}\bra{p_2=y}.
\end{align}
If we take into account the unitary balanced beam splitter, the projection state becomes
\begin{align}
\label{EQNProjectionByTwoHomodynesWithBalancedBeamSplitter}
\int dx_2 \Bra{\frac{q+x_2}{\sqrt{2}}}\Bra{\frac{-q+x_2}{\sqrt{2}}}e^{-iyx_2}.
\end{align}
During the measurement, this state will be jointly projected onto the measured and the ancillary state.
The measured state is unknown, but we can already apply the ancilla in the second mode.
This reduces the state to
\begin{align}
\label{EQNProjectionByHeterodyneWithAncillaryState}
\int dx \Bra{\frac{q+x}{\sqrt{2}}}\psiA\Parens{\frac{-q+x}{\sqrt{2}}}e^{-iyx},
\end{align}
where the subscript was dropped because it was no longer needed.
After a straightforward substitution we can express the projection state as
\begin{align}
\label{EQNExplicitPureHeterodyneProjector}
\int dx \psiA^\ast(x)e^{i \sqrt{2} x y}\ket{x + \sqrt{2} q}.
\end{align}
Since the time-reversal operator corresponds to complex conjugate in the $x$ representation, the expression~\eqref{EQNExplicitPureHeterodyneProjector} is the same as Eq.~\eqref{EQNPureHeterodyneProjector}\@.
For $q=y=0$, we obtain simple projection onto the given ancillary state $\int\psi^\ast(x)\ket{x}\,dx$.
We can see that if the ancillary mode is in the vacuum or a coherent state, the measurement remains the simple heterodyne detection, as is expected.
However, if the ancilla is non-Gaussian, we obtain a truly non-Gaussian measurement.

\subsection{Projecting on impure states}

In a realistic scenario, the ancillary state will be generally not pure.
To take this into account, it is best to abandon the $x$ representation and employ the formalism of Wigner functions.
The basic premise, however, remains.
The measurement still implements projection onto a specific state, only this time the state will be represented by a Wigner function.
Specifically, for a signal two mode state represented by a Wigner function $\WS(x_0, p_0, x_1, p_1)$, the outcome of a measurement performed on mode 1 yielding a pair of values $q$ and $y$ results in the Wigner function
\begin{align}
\begin{split}
&\Wout(x_0, p_0|q, y) \\
&\propto\int dx_1 dp_1 \WS(x_0, p_0, x_1, p_1)\WM(x_1, p_1 | q, y),
\end{split}
\end{align}
where the function $\WM(x_1, p_1 | q, y)$ represents the projector on the particular state.
In our scenario, in which the pair of homodyne detectors are supplied with an ancillary state corresponding to a Wigner function $\WA(x, p)$, the projector function can be found as
\begin{align}
\label{EQNImpureHeterodyneProjector}
\WM(x, p | q, y)=2\WA(x - \sqrt{2} q, -p + \sqrt{2}y).
\end{align}
We can see that this form agrees with Eq.~\eqref{EQNPureHeterodyneProjector} if we realize that the time-reversal operator $\hat{\mathcal{T}}$ transforms the Wigner function variables as $(x, p)\mapsto(x, -p)$.
The relation~\eqref{EQNImpureHeterodyneProjector} can be derived in the same way as relation~\eqref{EQNPureHeterodyneProjector}\@.
We start with the homodyne measurement projector functions, here represented by the pair of delta functions $\delta(x_1-q)\delta(p_2-y)$, which we then propagate through the beam splitter and apply to the ancillary state, resulting in
\begin{align}
\label{EQNWignerFunctionOfHeterodyneProjection}
\begin{split}
&\WM(x_1, p_1 | q, y) = \\
&\int dx_2 dp_2 \WA(x_2, p_2)\delta\Parens{\frac{x_1-x_2}{\sqrt{2}}-q}\delta\Parens{\frac{p_1+p_2}{\sqrt{2}}-y}.
\end{split}
\end{align}

\subsection{Arbitrary Gaussian operations within the measurement}
\label{SSCArbitraryGaussianOperationsWithinTheMeasurement}

One may desire to apply Gaussian operation to the non-Gaussian ancilla because some Gaussian operations (such as squeezing) enhance certain features of the state.
Here we show that, instead of projecting on a raw non-Gaussian state, we can alter the measurement so it projects on a non-Gaussian state altered by an arbitrary Gaussian operation.
This can be enormously useful because we do not need to implement an additional Hamiltonian that often makes the state impure in actual experiments.
Note that we are disregarding displacement because that can be achieved simply by displacing the measurement results.
For a pair of quadrature variables $x$ and $p$, an arbitrary Gaussian operation is represented by a real two-by-two symplectic matrix $S$ whose elements satisfy $s_{11}s_{22}-s_{12}s_{21}=1$.
If we consider that phase shift can be implemented ``for free,'' the arbitrary Gaussian unitary transformation reduces to
\begin{align}
x' = z_1x, \quad p' = \frac{1}{z_1}p + z_2x,
\end{align}
where $z_1$ and $z_2$ are arbitrary real parameters.
To achieve this transformation, we must modify the measurement setup in two ways.
First, the balanced beam splitter in Eq.~\eqref{EQNProjectionByTwoHomodynesWithBalancedBeamSplitter} will be removed and replaced by a beam splitter with transmittance $T$ and reflectance $R = 1 - T$.
Second, instead of measuring quadrature $p_2$ we measure quadrature $p_2(\theta) = p_2 \cos\theta + x_2\sin \theta$.
The projection functions of the measurements itself in Eq.~\eqref{EQNWignerFunctionOfHeterodyneProjection} are then
\begin{align}
\delta(x_1 - q) \delta(p_2 \cos\theta + x_2 \sin \theta - y).
\end{align}
Using the same steps we used to arrive at Eq.~\eqref{EQNImpureHeterodyneProjector} we can now obtain the generalized projection function
\small
\begin{align}
\label{EQNGeneralWignerFunctionOfHeterodyneProjection}
\begin{split}
& W_\mathrm{M}(x,p | q,y) = \frac{1}{|\sqrt{RT}\cos\theta|} \times \\
& W_\mathrm{A}\left(\sqrt{\frac{T}{R}}x - \frac{q}{\sqrt{R}}, -\sqrt{\frac{R}{T}} p - \frac{\tan \theta}{\sqrt{RT}}x + \frac{q \tan\theta}{\sqrt{R}}  + \frac{y}{\sqrt{T}\cos\theta}\right).
\end{split}
\end{align}
\normalsize
We can immediately see that after the time-reversal operations, we have $z_1 = \sqrt{T/R}$ and $z_2 = \tan\theta/\sqrt{RT}$ and these two parameters can attain arbitrary real values.
As a consequence, after addition of a phase shift the function~\eqref{EQNGeneralWignerFunctionOfHeterodyneProjection} implements projection onto the ancillary state altered by an arbitrary Gaussian operation.

It is worth pointing out that the two homodyne measurements need not to be independent.
One of the measurements can have parameters changing based on the results of the other one, thus creating a sort of adaptive measurement scheme.
For example, the measurement phase $\theta$ can depend on the measurement result $q$.
This can be used to induce a nonlinear behavior, as we will see in Sec.~\ref{SSCWithAdaptiveNonGaussianMeasurement}\@.

\section{Implementation of a cubic gate}
\subsection{With nonadaptive non-Gaussian measurement}
\label{SSCWithNonadaptiveNonGaussianMeasurement}

In this section we apply the non-Gaussian measurement to a particular task---the implementation of a nonlinear cubic gate $\hat{U} = e^{i \gamma \hat{x}^3}$ to an arbitrary quantum state.
In terms of quadrature operators, the gate performs transformation
\begin{align}
\label{EQNCubicOperators}
\hat{x}' = \hat{x}, \quad \hat{p}' = \hat{p} + 3\gamma\hat{x}^2.
\end{align}
Before proceeding to a scheme with the adaptive heterodyne measurement, we firstly consider implementation with nonadaptive measurement expressed by Eq.~\eqref{EQNPureHeterodyneProjector}\@.

The basic principle of the operation can be quickly explained in the $x$ representation.
The unknown input state $\ket{\psi}$ is mixed with a squeezed state on a balanced beam splitter.
If we for ease of explanation consider the infinite squeezing, the resulting two mode state can be expressed as
\begin{align}
\int dx \psi(x) \Ket{\frac{x}{\sqrt{2}}} \Ket{\frac{x}{\sqrt{2}}}.
\end{align}
After applying non-Gaussian measurement~\eqref{EQNProjectionByHeterodyneWithAncillaryState} on one of the modes, we obtain the projected state in the form
\begin{align}
\int dx \psi_\mathrm{A}\left(\frac{x}{\sqrt{2}} - \sqrt{2} q \right)\psi(x) e^{-i x y} \Ket{\frac{x}{\sqrt{2}}},
\end{align}
where $q$ and $y$ are again the homodyne measurement results, and $\psi_\mathrm{A}(x)$ is the wave function of the ancillary state.
For implementing the cubic gate, $\psi_\mathrm{A}(x)$ has to be cubically dependent on $x$, and is ideally in a state
\begin{align}
\label{EQNIdealCubicState}
\ket{\psi_\mathrm{A}} = \int dx \exp( i \gamma x^3) \ket{x}
\end{align}
and the whole operation would lead to
\begin{align}
\label{EQNIdealCubicGateBeforeFeedforward}
\begin{split}
& \exp(-i 3 \sqrt{2} \gamma q \hat{x}^2) \exp[i (6 \gamma q^2 - \sqrt{2} y) \hat{x}] \\
& \quad \times \exp(i \gamma \hat{x}^3) \int dx \psi(x) \Ket{\frac{x}{\sqrt{2}}}.
\end{split}
\end{align}
This is almost exactly the desired output state.
The only difference is a constant squeezing and two unitary operations depending on the measured values.
The constant squeezing can be fully compensated either before or after the operation and the measurement dependent unitary operations can be removed by a proper feedforward.
This is exactly the same principle as employed by the CV teleportation and CV measurement-induced operations.
While each particular measurement result projects on a different quantum state, these states belong to the same family and the proper operation can smear the differences and produce a quantum state independent of the measurement result.
This allows the whole procedure to operate deterministically.

\subsection{With adaptive non-Gaussian measurement}
\label{SSCWithAdaptiveNonGaussianMeasurement}

\begin{figure}
\centering
\includegraphics{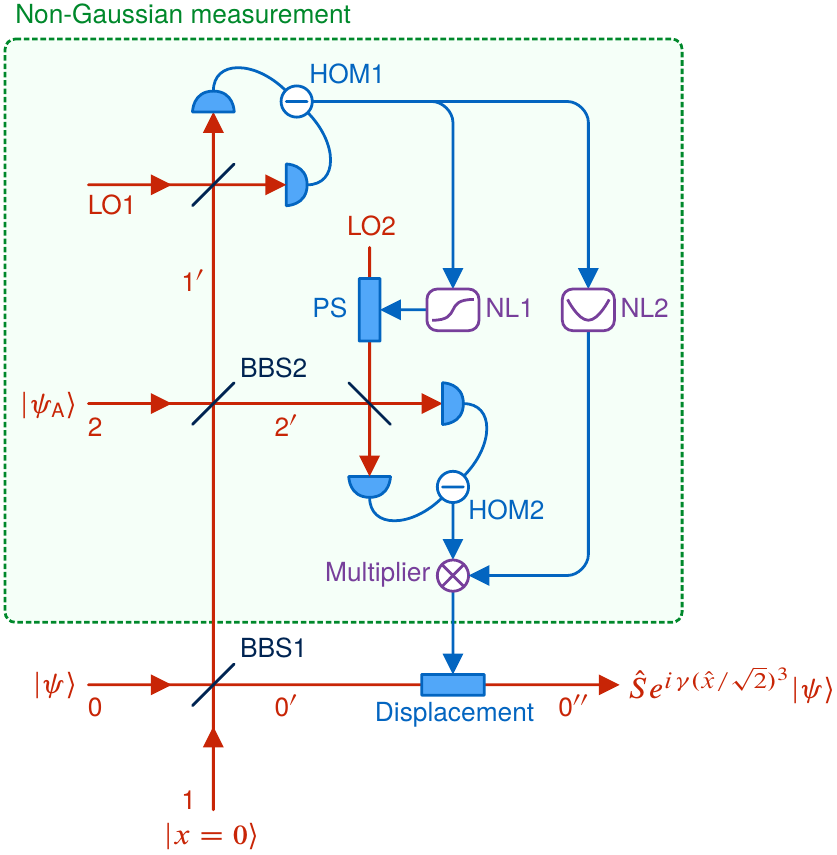}
\caption{(Color online)
Schematic of a cubic gate.
BBS, balanced beam splitter;
HOM, homodyne measurement;
LO, local oscillator;
PS, phase shift;
NL, nonlinear classical calculation.
While all the optics are linear, the classical circuit involves nonlinear calculations that makes the feedforward nonlinear.
The nonlinear classical circuits have been already devised in the experiment of dynamic squeezing \cite{Miyata2014}.
}
\label{FIGCPGSchematic}
\end{figure}

In Eq.~\eqref{EQNIdealCubicGateBeforeFeedforward} we need quadratic feedforward in the form of adjustable squeezing.
Thus the topology here is not as simple as the minimal implementation depicted in Fig.~\ref{FIGMinimalMBQCSchematic}\@.
To realize measurement of the nonlinear quadrature $\hat{p} + 3 \gamma \hat{x}^2$, we exploit the adaptive non-Gaussian heterodyne measurement.
According to the results in Sec.~\ref{SSCArbitraryGaussianOperationsWithinTheMeasurement}, by altering the phase of the second measurement, we can project onto a transformed ancillary state
\begin{align}
\hat{D}\Squares{\sqrt{2} q + i \Parens{\frac{\sqrt{2}y}{\cos \theta} - \sqrt{2} q \tan \theta}} \hat{\mathcal{T}} e^{i \hat{x}^2 \tan \theta} \ket{\psiA}.
\end{align}
Again, $q$ and $y$ are the measured values in the heterodyne detection,
and $\ket{\psiA}$ is the cubic state~\eqref{EQNIdealCubicState}\@.
We then substitute $3\sqrt{2} \gamma q$ for $\tan \theta$.
After simple algebras we find the projection state
\begin{align}
\exp(-i \gamma \hat{x}^3) \Ket{p = \frac{\sqrt{2} y}{\cos \theta}},
\end{align}
which means an eigenstate of the nonlinear quadrature $\hat{p} + 3 \gamma \hat{x}^2$ with the eigenvalue $\sqrt{2} y / \cos \theta$.
This scheme can be illustrated as Fig.~\ref{FIGCPGSchematic}\@.
Here the quadrature basis of the second homodyne detection is determined by the result of the first homodyne detection.
As a result, the heterodyne detection and the classical calculation compose a module of non-Gaussian measurement, and the feedforward is now the simple displacement operation.

To explicitly show how this scheme works, it is instructive to employ the Heisenberg representation, which would have the added benefit of incorporating the imperfections arising from the realistic experimental implementation, e.g.\@ finite squeezing.
 May the unknown signal mode be labeled by `0' and described by quadrature operators $\hat{x}_0$ and $\hat{p}_0$.
After combining the initial state in mode `0' with the squeezed state in mode `1' and with the non-Gaussian ancilla in mode `2', the respective quadrature operators read
\begin{subequations}
\label{EQNCubicGateAfterBBSMode0}
\begin{align}
\hat{x}_0'
  ={}& \frac{1}{\sqrt{2}} \hat{x}_0
     - \frac{1}{\sqrt{2}} \hat{x}_1, \\
\hat{p}_0'
  ={}& \frac{1}{\sqrt{2}} \hat{p}_0
     - \frac{1}{\sqrt{2}} \hat{p}_1,
\end{align}
\end{subequations}
\vspace{-\belowdisplayskip}
\begin{subequations}
\label{EQNCubicGateAfterBBSMode1}
\begin{align}
\hat{x}_1'
  ={}& \frac{1}{2} \hat{x}_0
     + \frac{1}{2} \hat{x}_1
     - \frac{1}{\sqrt{2}} \hat{x}_2, \\
\hat{p}_1'
  ={}& \frac{1}{2} \hat{p}_0
     + \frac{1}{2} \hat{p}_1
     - \frac{1}{\sqrt{2}} \hat{p}_2,
\end{align}
\end{subequations}
\vspace{-\belowdisplayskip}
\begin{subequations}
\label{EQNCubicGateAfterBBSMode2}
\begin{align}
\hat{x}_2'
  ={}& \frac{1}{2} \hat{x}_0
     + \frac{1}{2} \hat{x}_1
     + \frac{1}{\sqrt{2}} \hat{x}_2, \\
\hat{p}_2'
  ={}& \frac{1}{2} \hat{p}_0
     + \frac{1}{2} \hat{p}_1
     + \frac{1}{\sqrt{2}} \hat{p}_2.
\end{align}
\end{subequations}
In the next step we measure the $x$ quadrature of mode $1'$ and obtain value $q$.
We can now use the value to adjust the measured phase of the second homodyne detector.
In effect we end up measuring the value $y$ of quadrature operator $\hat{x}_2' \sin \theta + \hat{p}_2' \cos \theta$, where $\theta = \arctan(3 \sqrt{2} \gamma q)$.
Note that, since $\theta$ nonlinearly depends on $q$, which carries information of $\hat{x}_1$ quadrature, we can interpret this type of measurement as the origin of nonlinearity of the gate.
The quadrature operators of the output mode can be now expressed in terms of the measured values as
\begin{subequations}
\begin{align}
\hat{x}_0'
  ={}& \frac{1}{\sqrt{2}} \hat{x}_0
     - \frac{1}{\sqrt{2}} \hat{x}_1, \\
\hat{p}_0'
  ={}& \sqrt{2} \hat{p}_0 + p_2 +\frac{3\gamma}{2}[(\hat{x}_0 + \hat{x}_1)^2 - 2 \hat{x}_2^2] - \frac{\sqrt{2} y}{\cos\theta}
\end{align}
\end{subequations}
The last term of the $p$ quadrature, which is the only term explicitly depending on the measured values $q$ and $y$, can be removed by a suitable displacement and we are then left with the final form of the operators:
\begin{subequations}
\label{EQNCubicGateOutput}
\begin{align}
\label{EQNCubicGateOutputX}
\hat{x}_0''
  ={}& \frac{1}{\sqrt{2}} \hat{x}_0 - \frac{1}{\sqrt{2}} \hat{x}_1, \\
\label{EQNCubicGateOutputP}
\begin{split}
\hat{p}_0''
  ={}& \sqrt{2}
     \Parens{\hat{p}_0 + \frac{3 \gamma}{2 \sqrt{2}} \hat{x}_0^2} \\
  & + (\hat{p}_2 - 3 \gamma \hat{x}_2^2) + 3 \gamma \Parens{\hat{x}_0\hat{x}_1 + \frac{1}{2}\hat{x}_1^2}.
\end{split}
\end{align}
\end{subequations}
Both of the first terms in Eq.~\eqref{EQNCubicGateOutput} represent the ideal cubic operation, i.e.\@ combination of the cubic gate $e^{i \gamma (\hat{x}_0/\sqrt{2})^3}$ and the constant squeezing mentioned in Sec.~\ref{SECMinimalImplementationOfMeasurementInducedQuantumOperations}\@.
Those terms does not depend on the quadratures of the other ancillary states.
Differently from the output~\eqref{EQNIdealCubicGateBeforeFeedforward} in Sec.~\ref{SSCWithNonadaptiveNonGaussianMeasurement}, in the Heisenberg representation we can say that the cubic nonlinearity comes from the adaptive non-Gaussian measurement and feedforward process regardless of the ancillary states.

Naturally, the ancillary states are still required to complete the operation since the outputs have residual terms.
It is straightforward to find the ideal ancillary state in the mode `1' the quadrature eigenstate $\ket{x = 0}_1$ because the state affects only on the last terms of Eq.~\eqref{EQNCubicGateOutput} and they vanish when $\hat{x}_1 \to 0$.
In experimental implementation, we approach the ideal state by using squeezed vacuum states.

On the other hand, the middle term of Eq.~\eqref{EQNCubicGateOutputP}, $\hat{p}_\mathrm{NLQ} = \hat{p}_2 - 3 \gamma \hat{x}_2^2$, depends solely on the ancilla in the mode `2'.
This term vanishes when the ancilla is the ideal cubic state~\eqref{EQNIdealCubicState}\@.
This state is best approached by considering physical states that squeeze the nonlinear quadrature $\hat{p}_\mathrm{NLQ}$, as discussed in the next section.

\section{Optimum ancillary state}
\label{SECIdealAncillaryState}

To find suitable states in the mode `2', we can use the expectation value and the variance of the nonlinear quadrature $\hat{p}_\mathrm{NLQ}$ as figures of merit, both of which should be approaching zero.
Here we consider preparing the ancillary state that can be generated within reach of current technologies.
On one hand, arbitrary superpositions of photon-number states up to three photon level $\ket{\psi_{N = 3}}$ can be prepared \cite{Yukawa2013, Yukawa2013a}, and the photon-number limit can be in principle incremented.
On the other hand, we can perform universal Gaussian operation $\hat{U}_\mathrm{G}$ onto any input state \cite{Yoshikawa2007, Yoshikawa2008, Yoshikawa2011}.
Then the ancilla best suited for our purposes can be found in a form $\hat{U}_\mathrm{G} \ket{\psi_N}$ by optimizing over all superposition states up to $N$-photon level $\ket{\psi_N}$ and all Gaussian operations $\hat{U}_\mathrm{G}$ that can be applied on the state afterwards.
In this way, we are using the expensive non-Gaussian resources only for the key non-Gaussian features of the state \cite{Menzies2009}.

Our goal is to find a state $\hat{U}_\mathrm{G} \ket{\psi_N}$ that minimizes the expectation value $\moment{\hat{p}_\mathrm{NLQ}}$ and the variance $V(\hat{p}_\mathrm{NLQ}) = \moment{(\hat{p}_\mathrm{NLQ} - \moment{\hat{p}_\mathrm{NLQ}})^2}$.
The operator is symmetric with respect to space inversion, $\hat{x}_2 \rightarrow -\hat{x}_2$, and has a linear term of $\hat{p}_2$.
Accordingly the relevant Gaussian operations are the $p$ displacement represented by $\hat{p}_2 \to \hat{p}_2 + p_0$, and the $x$ squeezing represented by $\hat{x}_2 \to \hat{x}_2 / \lambda$ and $\hat{p}_2 \to \lambda \hat{p}_2$, where $p_0$ and $\lambda$ are arbitrary real parameters.
Thus the nonlinear quadrature after suitable Gaussian operations is represented as
\begin{align}
\hat{U}_G^\dagger \hat{p}_\mathrm{NLQ} \hat{U}_G = \gamma^{1/3} \Squares{\lambda' \hat{p}_2 - 3 \Parens{\frac{\hat{x_2}}{\lambda'}}^2} + p_0,
\end{align}
where $\lambda' = \lambda / \gamma^{1/3}$.
From this point of view, we can see that the expectation value $\moment{\hat{p}_\mathrm{NLQ}}$ vanishes when we apply suitable displacement $p_0$.
On the other hand, the variance $V(\hat{p}_\mathrm{NLQ})$ can be minimized by optimizing the state $\ket{\psi_N}$ and the parameter $\lambda'$.
Furthermore, since $\lambda'$ can be any real number, we can say that the optimum state does not depend on $\gamma$.
We therefore use the variance of $\lambda' \hat{p}_2 - 3 (\hat{x}_2 / \lambda')^2$ as the actual figure of merit to derive the optimum state $\ket{\psi_N}$ and the corresponding parameter $\lambda'$.

\begin{figure}
\centering
\includegraphics{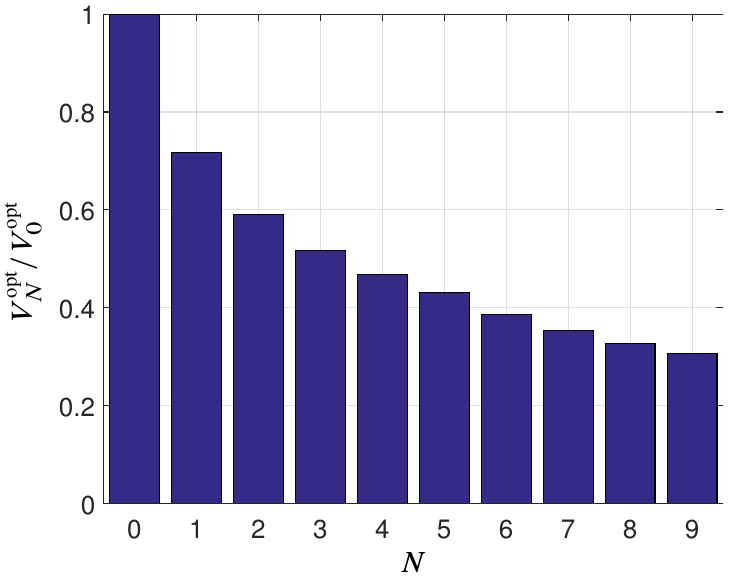}
\caption{(Color online)
Variances of the nonlinear quadrature with the optimized photon-number-state superpositions up to $N$ photons, normalized by the Gaussian limit $V_0^\mathrm{opt}$.
The parameter $\lambda'$ is optimized over to find the minimum of the variance.
}
\label{FIGNormalizedMinimumVarianceBars}
\end{figure}
\begin{figure}
\centering
\includegraphics{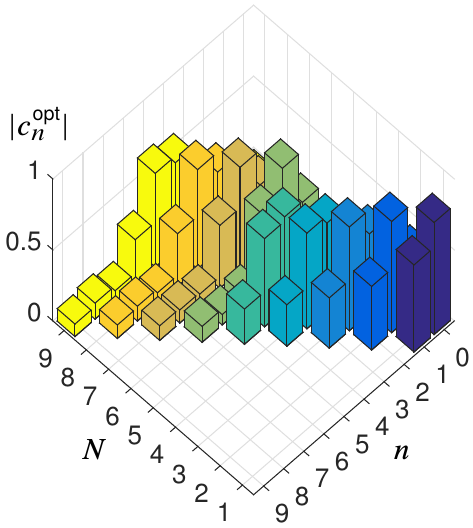}
\caption{(Color online)
Absolute values of the coefficients of the optimal finite approximation of ancillary states for various upper bounds of photon number $N$.
Note that the coefficients of even-number photons are real and odd-number photons imaginary, due to the symmetry of the nonlinear quadrature $\hat{p} - 3 \gamma \hat{x}^2$ with respect to $\hat{x} \to -\hat{x}$.
}
\label{FIGOptimizedCoefficients}
\end{figure}

Let $V_N^\mathrm{opt}$ the minimum value of the variance $V(\hat{p}_\mathrm{NLQ})$ with the optimum state $\ket{\psi_N^\mathrm{opt}}$ and the optimal parameter $\lambda'^{\mathrm{opt}}$.
Note that $V_0^\mathrm{opt}$ represents the Gaussian limit---the minimum variances when the state is optimized over all Gaussian states.
Then the relative noise $V_N^\mathrm{opt} / V_0^\mathrm{opt}$, as shown in Fig.~\ref{FIGNormalizedMinimumVarianceBars}, represents the ratio of the minimum noise to the Gaussian limit, and is independent from $\gamma$.
We can see that the variance decreases approaching zero with $N$, and that even a state obtained as a superposition of zero and one photon gives a substantial benefit over the Gaussian limit.
To present the optimized states, we represent the optimal approximate state up to $N$-photon level by $\ket{\psi_N^\mathrm{opt}} = \sum_{n=0}^N c_n^\mathrm{opt} \ket{n}$ and plot absolute values of the coefficients in Fig.~\ref{FIGOptimizedCoefficients}\@.
In the case of optimizing the superposition state up to three photons, the optimal approximate state looks as
\begin{align}
\label{EQNOptimizedStateUpTo3}
\ket{\psi_{N = 3}^\mathrm{opt}} \propto 0.17\ket{0} - 0.56i\ket{1} - 0.73\ket{2} + 0.35i\ket{3}.
\end{align}
The state is different from the cubic state from \cite{Marek2011} because of the different derivation of the states.
In \cite{Marek2011}, the state was determined as if it was produced by applying the cubic gate to the vacuum without considering optimization over squeezing and displacement.
On the other hand, the present state~\eqref{EQNOptimizedStateUpTo3} is derived so that its overall suitability as the ancilla is maximized with suitable Gaussian operations.
In either case, the state can be prepared by the same experimental method \cite{Yukawa2013, Yukawa2013a}.

\begin{figure}
\centering
\includegraphics{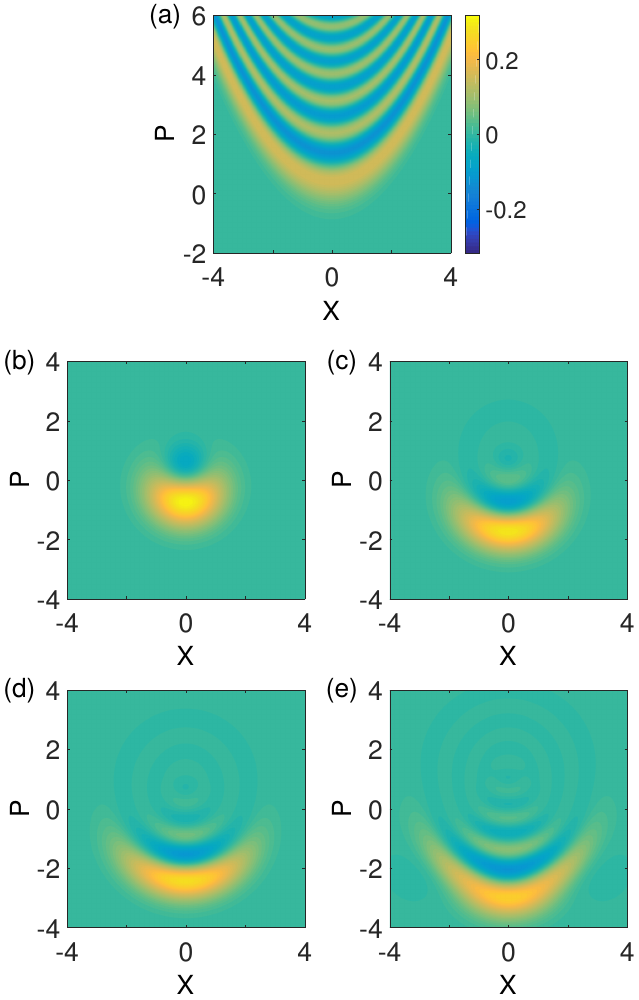}
\caption{(Color online)
Wigner functions of the optimal ancillary states.
(a) The ideal cubic state for $\gamma = 0.1$ (normalized over the displayed area), (b) $N = 1$, (c) $N = 3$, (d) $N = 5$, (e) $N = 9$.
Note that the approximate states have offsets in the $p$ direction, which can be compensated by $p$ displacement.
}
\label{FIGOptimizedStateWignerFunctions}
\end{figure}

The cubic nature of the states is also nicely visible from their Wigner functions as depicted in Fig.~\ref{FIGOptimizedStateWignerFunctions}\@.
For comparison, we check the Wigner function of the ideal cubic state \cite{Ghose2007}
\begin{align}
W(x, p) = 2 \pi \mathcal{N} \Abs{\frac{4}{3 \gamma}}^{1/3} \operatorname{Ai}\Parens{\Squares{\frac{4}{3 \gamma}}^{1/3} [3 \gamma x^2 - p]},
\end{align}
where $\operatorname{Ai}(x)$ is the Airy function and $\mathcal{N}$ a normalization factor.
As can be seen from Fig.~\ref{FIGOptimizedStateWignerFunctions} (a), the Wigner function is symmetric with respect to the $p$ axis and has a oscillating parabolic shape.
These characteristics also appear in the approximate cubic states shown in Fig.~\ref{FIGOptimizedStateWignerFunctions} (b)--(e).
As the upper limit of photon number becomes larger, the number of fringes along the $p$ direction increases approaching the ideal one.
Those Wigner functions of the approximate states can be considered to show core non-Gaussianity that then spreads out on the phase space by the following optimized squeezing.

So far we have not consider how to implement the optimized squeezing onto the core non-Gaussian state.
Actually, instead of adding another squeezing gate, the squeezing operation can be embedded into the adaptive non-Gaussian measurement by using the results in Sec.~\ref{SSCArbitraryGaussianOperationsWithinTheMeasurement}\@.
We discuss the details on it in Appendix~\ref{APPCubicGateWithUnbalancedAdaptiveNonGaussianMeasurement}\@.

\section{Conclusion}
We have introduced the concept of an adaptive non-Gaussian measurement---a CV measurement with a set of possible values, each of which is associated with a projection onto a non-Gaussian state.
The measurement is realized by a pair of homodyne detectors and a supply of suitable non-Gaussian ancillary states.
One particular advantage of this measurement is that an arbitrary Gaussian operation can be implemented on the soon to be measured quantum system simply by tools of passive linear optics.
In addition, some non-Gaussian operations can be implemented in a same way by making some of the measurement parameters dependent on already measured values.

To demonstrate this design feature, we have proposed a new method of realizing the cubic gate \cite{Marek2011}.
The current proposal does not require active operations to be performed on the transformed quantum system, all of them being part of the non-Gaussian measurement, which significantly improves the feasibility of the setup.
Specifically in the Heisenberg representation, it turns out that nonlinearity of the gate is created classically while the nonclassicality is given by the non-Gaussian ancilla in terms of reducing residual noise.
By exploiting the noise term as a figure of merit, we have found a new class of ancillary states that promise better performance than the states of \cite{Yukawa2013}.
The final implementation of the complete cubic gate can be therefore expected soon.

\begin{acknowledgments}
This work was partly supported by PDIS, GIA, APSA commissioned by the MEXT of Japan, ASCR-JSPS and the SCOPE program of the MIC of Japan.
P.M. and R.F. acknowledge a financial support from Grant No.\@ GA14-36681G of Czech Science Foundation.
H.Y. acknowledges the Australian Research Council, Grant No.\@ CECE1101027\@.
K.M. and H.O. acknowledge financial support from ALPS.
\end{acknowledgments}

\appendix

\section{Cubic gate with unbalanced adaptive non-Gaussian measurement}
\label{APPCubicGateWithUnbalancedAdaptiveNonGaussianMeasurement}

By replacing BBS2 in Fig.~\ref{FIGCPGSchematic} with an unbalanced beam splitter, we have another degree of freedom to effectively apply arbitrary squeezing operation onto the ancillary non-Gaussian state, as shown in Sec.~\ref{SSCArbitraryGaussianOperationsWithinTheMeasurement}\@.
Thus the Gaussian optimization discussed in Sec.~\ref{SECIdealAncillaryState} can be embedded in the cubic-gate schematic.

We explain here an input-output relationship of the cubic gate with unbalanced beam splitters.
Transmittance and reflectance of the first beam splitter are represented as $T_1$ and $R_1 = 1 - T_1$, respectively.
$T_2$ and $R_2$ are also defined in the same way for the second beam splitter.
After the beam splitter transformations, the quadratures of the output modes are
\begin{subequations}
\label{EQNGeneralizedCubicGateAfterBSMode0}
\begin{align}
\hat{x}_0'
  ={}& \sqrt{T_1} \hat{x}_0
     - \sqrt{R_2} \hat{x}_1, \\
\hat{p}_0'
  ={}& \sqrt{T_1} \hat{p}_0
     - \sqrt{R_2} \hat{p}_1,
\end{align}
\end{subequations}
\vspace{-\belowdisplayskip}
\begin{subequations}
\label{EQNGeneralizedCubicGateAfterBBSMode1}
\begin{align}
\hat{x}_1'
  ={}& \sqrt{R_1 T_2} \hat{x}_0
     + \sqrt{T_1 T_2} \hat{x}_1
     - \sqrt{R_2} \hat{x}_2, \\
\hat{p}_1'
  ={}& \sqrt{R_1 T_2} \hat{p}_0
     + \sqrt{T_1 T_2} \hat{p}_1
     - \sqrt{R_2} \hat{p}_2,
\end{align}
\end{subequations}
\vspace{-\belowdisplayskip}
\begin{subequations}
\label{EQNGeneralizedCubicGateAfterBBSMode2}
\begin{align}
\hat{x}_2'
  ={}& \sqrt{R_1 R_2} \hat{x}_0
     + \sqrt{T_1 R_2} \hat{x}_1
     + \sqrt{T_2} \hat{x}_2, \\
\hat{p}_2'
  ={}& \sqrt{R_1 R_2} \hat{p}_0
     + \sqrt{T_1 R_2} \hat{p}_1
     + \sqrt{T_2} \hat{p}_2.
\end{align}
\end{subequations}
After measuring the $x$ quadrature of mode $1'$ and obtain value $q$, we set the phase factor
\begin{align}
\theta = \arctan\Parens{\frac{6 T_2 \gamma}{\sqrt{R_2}} q}.
\end{align}
Then we measure the quadrature $\hat{x}_2' \sin\theta + \hat{p}_2' \cos\theta$ and obtain value $y$.
The $p$ quadrature of the unmeasured mode $0'$ can be expressed with the measured values $q$ and $y$ as
\begin{align}
\begin{split}
\hat{p}_0' ={}& \frac{1}{\sqrt{T_1}} \hat{p}_0 - \frac{\sqrt{R_1}}{\sqrt{T_1 R_2} \cos\theta} y \\
& + \sqrt{\frac{R_1 T_2}{T_1 R_2}} \hat{p}_2 - 6 \gamma \sqrt{\frac{R_1}{T_1}} \Parens{\frac{T_2}{R_2}}^{3/2} q^2 \\
& + \Parens{\frac{6 R_1 T_2 \gamma}{\sqrt{T_1} R_2^{3/2}} \hat{x}_0 + \frac{6 \sqrt{R_1} T_2 \gamma}{R_2^{3/2}} \hat{x}_1} q.
\end{split}
\end{align}
We apply $p$ displacement to this quadrature with value
\begin{align}
\label{EQNDisplacementInGeneralizedCPG}
p_\mathrm{disp} = \frac{\sqrt{R_1}}{\sqrt{T_1 R_2} \cos\theta} y + \frac{3\gamma \sqrt{R_1 T_2} (T_2 - R_2)}{\sqrt{T_1} R_2^{3/2}} q^2
\end{align}
and obtain the output quadratures
\begin{subequations}
\label{EQNGeneralizedCPGOutput}
\begin{align}
\label{EQNGeneralizedCPGOutputX} 
\hat{x}_0'' ={}& \sqrt{T_1} \Parens{\hat{x}_0 - \sqrt{\frac{R_1}{T_1}} \hat{x}_1}, \\
\label{EQNGeneralizedCPGOutputP}
\begin{split}
\hat{p}_0'' ={}& \frac{1}{\sqrt{T_1}} \left\{\Squares{\hat{p}_0 + 3 \gamma \Parens{\frac{R_1 T_2}{R_2}}^{3/2} \hat{x}_0^2} \right. \\
& + \sqrt{\frac{R_1 T_2}{R_2}} (\hat{p}_2 - 3 \gamma \hat{x}_2^2) \\
& + \left. 6 \gamma R_1 \sqrt{T_1} \Parens{\frac{T_2}{R_2}}^{3/2} \Parens{\hat{x}_0 \hat{x}_1 + \frac{1}{2} \sqrt{\frac{T_1}{R_1}} \hat{x}_1^2} \right\}.
\end{split}
\end{align}
\end{subequations}
We can see that the outputs are equal to Eq.~\eqref{EQNCubicGateOutput} if we set $T_1 = R_1 = T_2 = R_2 = 1/2$.
Note that, if we use unbalanced beam splitters, the displacement has a quadratic term as shown in Eq.~\eqref{EQNDisplacementInGeneralizedCPG}.

To explicitly see how the transmittances of the beam splitters affect on the quadratures of the ancillary non-Gaussian state, we scale the strength of cubic nonlinearity $\gamma$ to $(R_2 / R_1 T_2)^{3/2} \gamma$.
The output $p$ quadrature~\eqref{EQNGeneralizedCPGOutputP} is then expressed as
\begin{align}
\label{EQNGeneralizedCPGOutputPScaled}
\begin{split}
\hat{p}_0'' ={}& \frac{1}{\sqrt{T_1}} \left\{(\hat{p}_0 + 3 \gamma \hat{x}_0^2) \vphantom{\sqrt{\frac{T}{R}}} \right. \\
& + \Squares{\sqrt{\frac{R_1 T_2}{R_2}} \hat{p}_2 - 3 \gamma \Parens{\sqrt{\frac{R_2}{R_1 T_2}} \hat{x}_2}^2} \\
& + \left. 6 \gamma \sqrt{\frac{T_1}{R_1}} \Parens{\hat{x}_0 \hat{x}_1 + \frac{1}{2} \sqrt{\frac{T_1}{R_1}} \hat{x}_1^2} \right\}.
\end{split}
\end{align}
The second term represents the nonlinear noise determined by the non-Gaussian measurement.
We can see that the ancilla is effectively squeezed by the squeezing factor $\sqrt{R_1 T_2 / R_2}$, which can be fully controlled by choosing transmittance of the second beam splitter.
While universal squeezing operation in actual experiments \cite{Yoshikawa2007,Miwa2014,Miyata2014} adds nonnegligible noise to the input state because of finite squeezing in its resource state, the effective squeezing in the heterodyne measurement does not require additional resource states, which helps preparation of the approximate cubic state with high purity.

\section{Numerical method of approximating photon-number superposition to the cubic state}

In Sec.~\ref{SECIdealAncillaryState}, we considered the variance $V(\hat{p}_\text{NLQ})$ as a figure of merit to approximate the cubic state with photon-number-superposition states up to certain photon level and squeezing.
Intuitively, the approximation can be done by numerically optimizing all of the coefficients of a superposition state and the squeezing level, but it often leads to locally optimum solutions, especially when increasing the upper limit of photon numbers.
Here we reduce the problem into optimization with two variables, regardless of the size of the Hilbert space.
With each set of the two variables, an optimized superposition state can be derived as an eigenstate of the minimum eigenvalue of a certain positive-semidefinite operator.
By numerically creating a minimum-search map with the two variables, we can make sure that the solution is almost certainly the true optimum one.
The method is a variation of the classical variance-minimization problem \cite{Cai1993}.

Suppose $\mathcal{H}_N$ is a $(N+1)$-dimensional Hilbert space up to $N$-photon level, and $\ket{\psi}$ is a state in $\mathcal{H}_N$.
Our purpose is to find a set of the optimum state $\ket{\psi}$ and the optimum parameter $\lambda'$ that minimizes the variance of the nonlinear quadrature $\hat{y}(\lambda') = \lambda' \hat{p} - 3 (\hat{x} / \lambda')^2$.
This problem can be written as
\begin{subequations}
\begin{align}
& \min_{\substack{\ket{\psi} \in \mathcal{H}_N \\ \lambda' \in \mathbb{R}}} V(\ket{\psi}, \lambda'), \\
\label{EQNEvaluationVariance}
& V(\ket{\psi}, \lambda') = \braket{\psi}{[\hat{y}(\lambda') - \moment{\hat{y}(\lambda')}]^2}{\psi},
\end{align}
\end{subequations}
where $\moment{\hat{y}(\lambda')}$ is the expectation value $\braket{\psi}{\hat{y}(\lambda')}{\psi}$.

To make this problem digestible, we alternatively consider another minimization problem.
Let $d$ be a real number.
We then replace the expectation value in Eq.~\eqref{EQNEvaluationVariance} with $d$ and set a new evaluation function
\begin{align}
Z(\ket{\psi}, \lambda', d) = \braket{\psi}{[\hat{y}(\lambda') - d]^2}{\psi}.
\end{align}
Next, we introduce another evaluation function $W(d)$ defined as minimum of $Z(\ket{\psi}, \lambda', d)$ with respect to $\ket{\psi} \in \mathcal{H}_N$ and $\lambda' \in \mathbb{R}$.
This can be expressed as
\begin{align}
W(d) = \min_{\substack{\ket{\psi} \in \mathcal{H}_N \\ \lambda' \in \mathbb{R}}} Z(\ket{\psi}, \lambda', d).
\end{align}
Suppose $W(d)$ is minimum when $d = d^\star$.
In addition, suppose $Z(\ket{\psi}, \lambda', d^\star)$ is minimum when $\ket{\psi} = \ket{\psi^\star}$ and $\lambda' = \lambda'^\star$.
Then we can say that the set $(\ket{\psi^\star}, \lambda'^\star)$ is the true optimum set that minimizes $V(\ket{\psi}, \lambda')$.
This is verified as follows.
Let $\moment{\hat{y}(\lambda')}^\star$ be the expectation value $\braket{\psi^\star}{\hat{y}(\lambda')}{\psi^\star}$.
Then
\begin{align}
\begin{split}
Z(\ket{\psi^\star}, \lambda'^\star, d^\star) \leq{}& W(\moment{\hat{y}(\lambda'^\star)}^\star) \\
\leq{}& Z(\ket{\psi^\star}, \lambda'^\star, \moment{\hat{y}(\lambda'^\star)}^\star)
\end{split}
\end{align}
and consequently $(\moment{\hat{y}(\lambda'^\star)}^\star - d^\star)^2 \leq 0$, which means $d^\star = \moment{\hat{y}(\lambda'^\star)}^\star$. Therefore for any $\ket{\psi} \in \mathcal{H}_N$, any $\lambda' \in \mathbb{R}$ and the corresponding expectation value $\moment{\hat{y}(\lambda')} = \braket{\psi}{\hat{y}(\lambda')}{\psi}$, it holds that
\begin{align}
\begin{split}
& V(\ket{\psi^\star}, \lambda'^\star) = W(d^\star) \\
& \leq W(\moment{\hat{y}(\lambda')}) \leq{} Z(\ket{\psi}, \lambda', \moment{\hat{y}(\lambda')}) = V(\ket{\psi}, \lambda'),
\end{split}
\end{align}
which means $V(\ket{\psi^\star}, \lambda'^\star)$ is minimum.
As a result, the problem can be solved by searching for a state that minimizes $Z(\ket{\psi}, \lambda', d)$ with every $\lambda'$ and $d$.

The point is that $Z(\ket{\psi}, \lambda', d)$ is a quadratic form, and therefore each optimum state is determined as an eigenstate of the minimum eigenvalue of $[\hat{y}(\lambda') - d]^2$ represented by the limited Hilbert space.
In the case that we look for the optimum state up to $N$-photon level, the matrix representation of $[\hat{y}(\lambda') - d]^2$ reads
\begin{align}
Y(\lambda', d) ={}& \sum_{m, n = 0}^N Y_{m n}(\lambda', d) \ket{m} \bra{n}, \\
Y_{m n}(\lambda', d) ={}& \braket{m}{[\hat{y}(\lambda') - d]^2}{n},
\end{align}
and the optimum state in terms of $(\lambda', d)$ is found as the eigenstate of the minimum eigenvalue of the matrix $Y(\lambda', d)$, which can be deterministically obtained by numerical calculation.
This implies that the problem is now broken down into a two-variable optimization problem.
We can create a minimum-search map $\min_{\ket{\psi} \in \mathcal{H}_N} Z(\ket{\psi}, \lambda', d)$ with respect to $\lambda'$ and $d$, which makes it easy to look for the true optimum solution.

\begin{figure}
\centering
\includegraphics{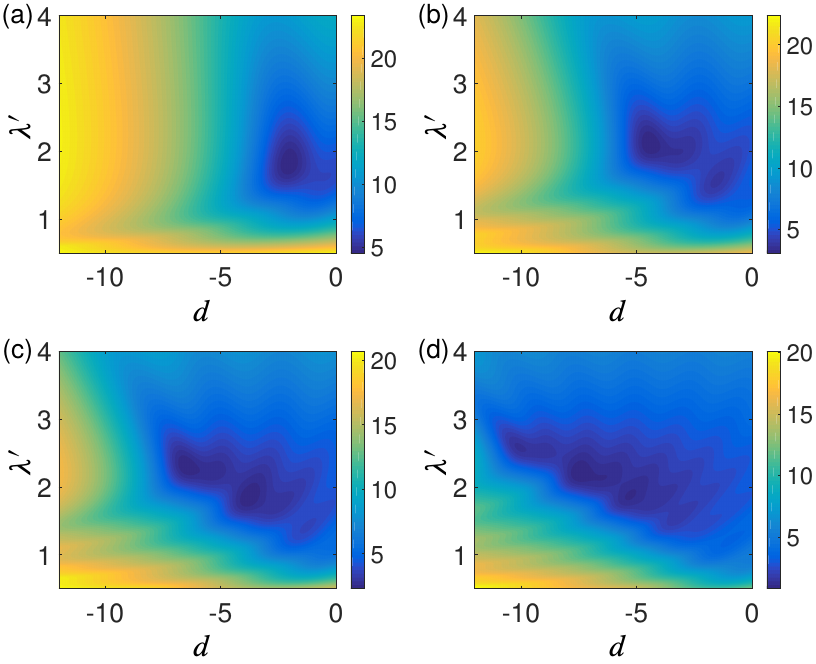}
\caption{(Color online)
Minimum-search map. (a) $N=1$, (b) $N=3$, (c) $N=5$, (d) $N=9$. The values are normalized by shot noise level and shown in dB scale.
}
\label{FIGMinimumSearchMap}
\end{figure}

Figure~\ref{FIGMinimumSearchMap} shows some examples of the map used to derive the optimized superposition states in Fig.~\ref{FIGOptimizedStateWignerFunctions}\@.
We can see that the number of local minima increases as the upper limit of photon numbers becomes larger.
By choosing suitable ranges and resolutions of $(\lambda', d)$, we almost certainly find the true minimum and consequently the true optimized state.

\hbadness = 10000
\bibliography{CPGScheme}

\end{document}